\documentclass[pra,aps,floatfix,showpacs,tightenlines,superscriptaddress,amsmath,amssymb,showkeys,10pt,nofootinbib]{revtex4-2}
\usepackage{slashed}
\usepackage{xcolor}
\usepackage{graphicx}

\usepackage{hyperref}

\newcommand{\be}{\begin{equation}}\newcommand{\ee}{\end{equation}}
\newcommand{\bea}{\begin{eqnarray}}\newcommand{\eea}{\end{eqnarray}}
\newcommand{\brr}{\begin{array}}\newcommand{\err}{\end{array}}
\newcommand{\bit}{\begin{itemize}}\newcommand{\eit}{\end{itemize}}
\newcommand{\ben}{\begin{enumerate}}\newcommand{\een}{\end{enumerate}}

\newcommand{\bbm}{\begin{bmatrix}}\newcommand{\ebm}{\end{bmatrix}}
\newcommand{\ba}{\begin{array}}
\newcommand{\ea}{\end{array}}
\newcommand{\G}{\textbf}

\newtheorem{mydef}{Definition}
\newtheorem{Lemma}{Lemma}
\newcommand{\bd}{\begin{mydef}} \newcommand{\ed}{\end{mydef}}
\newcommand{\bthe}{\begin{theorem}} \newcommand{\ethe}{\end{theorem}}
\newcommand{\ble}{\begin{Lemma}} \newcommand{\ele}{\end{Lemma}}

\newcommand{\dr}{\mathrm{d}}

\def\ha{\frac{1}{2}}

\def\intx{\int \!\!\mathrm{d}^3 {\G x}}

\def\lab{\label}\def\lan{\langle}
\def\lf{\left}

\def\non{\nonumber}\def\pa{\partial}\def\ran{\rangle}

\def\ri{\right}
\def\al{\alpha}\def\bt{\beta}\def\ga{\gamma}\def\Ga{\Gamma}
\def\de{\delta}\def\De{\Delta}

\def\si{\sigma}
\def\om{\omega}

\def\1{{_{1}}}\def\2{{_{2}}}

\newcommand{\ide}{1\hspace{-1mm}{\rm I}}

\def\noHe0{:\;\!\!\;\!\!:H_e(0):\;\!\!\;\!\!:}
\def\noHm0{:\;\!\!\;\!\!:H_\mu(0):\;\!\!\;\!\!:}

\def\lab{\label}
\def\lan{\langle}
\def\lf{\left}

\def\non{\nonumber}
\def\pa{\partial}\def\ran{\rangle}

\def\ri{\right}

\def\al{\alpha}\def\bt{\beta}\def\ga{\gamma}
\def\Ga{\Gamma}\def\de{\delta}\def\De{\Delta}

\def\si{\sigma}
\def\om{\omega}

\def\1{{_{1}}}\def\2{{_{2}}}

\begin{document}

\title{Neutrino oscillations in the interaction picture}

\author{Massimo Blasone}
\email{blasone@sa.infn.it}
\affiliation{Dipartimento di Fisica, Universit\`a di Salerno, Via Giovanni Paolo II 132, 84084 Fisciano (SA), Italy}
\affiliation{INFN Sezione di Napoli, Gruppo collegato di Salerno, Italy}

\author{Francesco Giacosa}
\email{francesco.giacosa@gmail.com}
\affiliation{Institute of Physics, Jan-Kochanowski University, ul. Uniwersytecka 7, 25-406 Kielce, Poland}
\affiliation{Institute for Theoretical Physics, J. W. Goethe University, Max-von-Laue-Straße 1,
60438 Frankfurt, Germany}

\author{Luca Smaldone}
\email{lsmaldone@unisa.it}
\affiliation{Dipartimento di Fisica, Universit\`a di Salerno, Via Giovanni Paolo II 132, 84084 Fisciano (SA), Italy}
\affiliation{INFN Sezione di Napoli, Gruppo collegato di Salerno, Italy}
%\affiliation{Faculty of Physics, University of Warsaw, ul. Pasteura 5, 02-093 Warsaw, Poland}

\author{Giorgio Torrieri}
\email{torrieri@unicamp.br}
\affiliation{Instituto de Fisica Gleb Wataghin - UNICAMP, 13083-859, Campinas SP, Brazil}
\affiliation{Institute of Physics, Jan-Kochanowski University, ul. Uniwersytecka 7, 25-406 Kielce, Poland}

\begin{abstract}
We study the mixing of different kind of fields (scalar in 0+1D, scalar in 3+1D, fermion in 3+1D) treating the mixing term as an interaction. To this aim, we employ the usual perturbative series in the interaction picture. We find that expression for flavor changing probability exhibits corrections with respect to the usual quantum mechanical (e.g. neutrino) oscillation formula, in agreement with the result previously obtained in the non-perturbative flavor Fock space approach.
\end{abstract}

\maketitle
%%%%%%%%%%%%%%%%%%%%%%%%%%%%%%%%%%%%%%%%%%%%%%%%%%%%%%%%%%%%%%%%%%%%%%%%%%%%%%%
\section{Introduction}

The idea of neutrino oscillations as a mechanism to solve the solar-neutrino puzzle was firstly proposed by Pontecorvo and collaborators \cite{Gribov:1968kq,Bilenky:1975tb,Bilenky:1976yj,Bilenky:1977ne} and it was later confirmed by a plethora of experiments (see e.g. \cite{Vogel:2015wua,IceCube:2017lak,Nakano:2020lol,OPERA:2021xtu}). 

Although many features of neutrino mixing and oscillations are now well-understood \cite{Strumia:2006db,giunti2007fundamentals,Fantini:2018itu}, there is no agreement on their correct ultimate description within quantum field theory (QFT). 
Various ideas were proposed in the last three decades, as \emph{external wavepackets} \cite{PhysRevD.48.4310,Beuthe:2001rc}, \emph{weak-process states} \cite{PhysRevD.45.2414} and the flavor Fock-space approach \cite{Blasone:1995zc, PhysRevD.59.113003, Hannabuss:2000hy, PhysRevD.64.013011, PhysRevD.65.096015, Hannabuss:2002cv,Lee:2017cqf}. The latter is based on the discovery \cite{Blasone:1995zc} that the flavor and the mass representations of the equal-time anticommutation relations of neutrino fields, are unitarily inequivalent \cite{friedrichs1953mathematical,barton1963introduction,berezin1966method,umezawa1993advanced}. Therefore, the Hilbert space where flavor fields are defined is explicitly built in and the oscillation probability is computed by taking the expectation value of lepton currents/charges on the one-particle neutrino states at a reference time. Such modified formula differs from the classic Pontecorvo result in two respects \cite{BHV99}: 
i) apart from the usual oscillation term which depends on the \emph{difference} of neutrino energies/frequencies, the oscillation formula of Ref.\cite{BHV99} shows up a fast-oscillation term which depends on the \emph{sum} of the frequencies;
ii) in the formula of Ref.\cite{BHV99}, there are energy dependent oscillation amplitudes which are the coefficients of a Bogoliubov transformation \cite{Blasone:1995zc}.

In this paper we introduce a different approach, in a close analogy to what is done in the study of unstable particles \cite{PhysRevLett.71.2687,facchi1999regola}. In fact, we employ the interaction (Dirac) picture, where the interaction Lagrangian in the Dyson series only contains the mixing term between different flavor-fields. For simplicity, we  limit our calculation to the case of two flavors. Then we compute amplitudes for the various decay channels at the first order, which describe both flavor changing and survival processes. Three examples are here analyzed: a quantum mechanical (QFT in 0+1D) toy model, a scalar field model and a fermion (``neutrino'') model in 3+1D. Remarkably, we find the that the fermion flavor-transition formula non-trivially agrees, within the approximation adopted, with the non-perturbative formula of the flavor-Fock space approach. Let us remark that the comparison is not possible in the boson case, where the flavor charge expectation value is not positive-definite and thus it cannot be interpreted as a probability \cite{Blasone:2001du,bigs2}. In this respect, the present work represents also a viable approach to compute the oscillation probability in this tricky situation.

The paper is organized as follows: in Section \ref{gencon} we present general considerations on field mixing and the interaction picture approach. In Section \ref{childsec} we study the 0+1D toy model, while in Sections \ref{bosesec} and \ref{neutsec} we extend our consideration to 3+1D scalar and fermion models, respectively. Finally, we present discussion and conclusions in Section \ref{con}. For reader's convenience, in Appendix \ref{comp} we briefly review the non-perturbative flavor Fock space approach.
%%%%%%%%%%%%%%%%%%%%%%%%%%%%%%%%%%%%%%%%%
\section{General considerations} \label{gencon}
The charged-current lepton sector of weak interaction is described (in the case of two-flavors) by the Lagrangian
\bea \lab{neutr}
\mathcal{L} \  =  \sum_{\si=e,\mu} \lf[\overline{\nu}_\si \lf(i \slashed{\pa}-m_\si\ri) \nu_\si+\overline{l}_\si \lf( i \ga_\mu \pa^\mu - \tilde{m}_\si \ri) l_\si \ri]  + \mathcal{L}_{mix}+\mathcal{L}_{wint} \, ,
\label{Linteract}
\eea
with
\bea
 \mathcal{L}_{mix} & = & -m_{e \mu} \lf(\overline{\nu}_e \nu_\mu+\overline{\nu}_\mu \nu_e\ri) \, , \\[2mm]
\mathcal{L}_{wint} & = & -\frac{g}{2\sqrt{2}}\sum_{\si=e,\mu}\lf [ W_{\mu}^{+}\, \overline{\nu}_\si\,\gamma^{\mu}\,(1-\gamma^{5})\,l_\si +
h.c. \ri]
\eea
The neutrino kinetic term (including $\mathcal{L}_{mix}$) can be diagonalized by the \emph{mixing transformation} \cite{Bilenky:1978nj,Bilenky:1987ty}  
\be \label{mixtra}
\nu_\si \ = \ \sum_{j=1,2} U_{\si j}^* \nu_j \, , 
\ee
$U$ is the \emph{mixing matrix}. In the two flavor case, here analyzed
\be
U \ = \ \begin{pmatrix} \cos \theta & \sin \theta \\ -\sin \theta & \cos \theta \end{pmatrix} \, , 
\ee
with $\tan 2 \theta = 2 m_{e \mu}/(m_\mu-m_e)$. 

If one employs the interaction picture to compute transition amplitudes, $\mathcal{L}$ must be decomposed into a free and an interaction part. A possible choice is  
\be
\mathcal{L} \ = \ \mathcal{L}_0^m+\mathcal{L}_{int}^m \, , 
\ee
with 
\bea
{\cal L}_0^m & =  & \sum_j \, \overline{\nu}_j  \lf( i \ga_\mu \pa^\mu - m_j \ri)\nu_j  \, + \, \sum_\si \, \overline{l} \lf( i \ga_\mu \pa^\mu - \tilde{m}_{\si} \ri) l \, ,  \\[2mm]
{\cal L}_{int}^m & = &  -\frac{g}{2\sqrt{2}} \sum_{\si,j} \,\lf[ W_{\mu}^{+}\, \overline{\nu}_{j} \, U^*_{j \si}\,\gamma^{\mu}\,(1-\gamma^{5})\, l_\si +
h.c. \ri] \, .
\eea
In such a case the effect of mixing is incorporated in the weak-interaction vertex. Following this approach, one is led to calculate transition amplitudes in which neutrinos appear only as internal lines \cite{PhysRevD.48.4310,Beuthe:2001rc,Naumov:2020yyv} \footnote{Note that, in this approach, the issue of inequivalence of flavor and mass vacua for neutrinos (see Appendix) is not taken into account.}.

However, in charged current weak interaction processes,  neutrinos are produced with a definite flavor. Therefore, another reasonable possibility is to take the following split
\be
\mathcal{L} \ = \ \mathcal{L}_0+\mathcal{L}^g_{int} \, , 
\ee
with
\bea \lab{L0}
\mathcal{L}_0 & = & \sum_{\si=e,\mu} \overline{\nu}_\si \lf(i \slashed{\pa}-m_\si\ri) \nu_\si+\sum_{\si=e,\mu} \overline{l}_\si \lf(i \slashed{\pa}-\tilde{m}_\si\ri) l_\si \, ,  \\[2mm]
\mathcal{L}^g_{int} & = & \mathcal{L}_{mix}+\mathcal{L}_{wint} \, . 
\eea
In this approach, $\mathcal{L}_{wint}$ is diagonal in the asymptotic fields appearing in Eq.\eqref{L0}. Thus, in order to describe neutrino oscillations, we can safely disregard $\mathcal{L}_{wint}$ (zeroth-order in $g$), so that the charged-lepton part also decouples. In other words, we can treat the mixing term as an interaction, and we can compute the transition amplitudes among different flavors by means of the usual Dyson formula for the time evolution operator
\be \label{dyfor}
U(t_i,t_f) \ = \ \mathcal{T} \exp \lf[i \int^{t_f}_{t_i} \!\! \dr^4 x \, :\mathcal{L}_{int}(x): \ri] \ = \ \mathcal{T}\exp  \lf[-i \int^{t_f}_{t_i} \!\! \dr^4 x \, :\mathcal{H}_{int}(x): \ri] \, ,
\ee
where $\mathcal{L}_{int} \equiv \mathcal{L}^{g=0}_{int}=-m_{e \mu} \lf(\overline{\nu}_e \nu_\mu+\overline{\nu}_\mu \nu_e\ri)$, $\mathcal{H}_{int}(x)=-\mathcal{L}_{int}(x)$ is the interaction Hamiltonian density and $\mathcal{T}$ is the chronological product. In the following we will only need the expression of the operator up to the second order
\be \label{dyfor2}
U(t_i,t_f) 
\ = \ 1-i\int_{t_{i}}^{t_{f}}\!\!\dr t_{1} \,H_{int}(t_{1})+(-i)^{2}%
\int_{t_{i}}^{t_{f}}\!\!\dr t_{1} \,H_{int}(t_{1})%
\int_{t_{i}}^{t_{1}}\dr t_{2} \, H_{int}(t_{2})+...
\ee
where $H_{int}=\intx \,\mathcal{H}_{int}(x)$ is the interaction Hamiltonian.

Notice that we look at the time evolution operator and not at the $S$-matrix. This is because the phenomenon of flavor oscillations can only be described at finite time. This amounts to say that flavor neutrino states do not exist as asymptotically stable states. As it will be clear from the various examples below, the limits $t_i \to - \infty$  and $t_f \to + \infty$ forbid the flavor-changing processes under study. At the same time, such a limit guarantees strict  energy conservation. This is in agreement with the \emph{flavor-energy uncertainty relation} derived in Ref. \cite{Blasone2019} and it is analogous to what happens for unstable particles \cite{PhysRevLett.71.2687,facchi1999regola,Giacosa:2010br,Giacosa:2011xa,Giacosa:2018dzm,Giacosa:2021hgl} (see also \cite{Anselmi:2023wjx,Anselmi:2023phm}, where the importance of finite-time QFT in the study of decay has been emphasized). As a matter of fact, both the decay of unstable particles \cite{Bhattacharyya_1983} and neutrino oscillations \cite{Bilenky:2009zz} can be viewed in terms of the \emph{time-energy uncertainty relations}. 

In the following we will first study the case of 0+1D QFT (that is, QM), and a 3+1D scalar model. This preliminary analysis permits to grasp the main features of the problem, without the complication of dealing with spinors.
%%%%%%%%%%%%%%%%%%%%%%%%%%%%%%%%%%%%%%%%%%%%%%%%%%%%%%%%%%%%%%%%%%%
\section{A quantum mechanics toy model of flavor mixing} \label{childsec}

Let us consider the quantum mechanical problem of two interacting harmonic
oscillators with bare frequencies $\om_{A,B}$. 
We treat this problem as a $0+1D$ field theory described by the Lagrangian 
\begin{equation}
L=\frac{1}{2}\left( \frac{dx_{A}}{dt}\right) ^{2}-\frac{\om_{A}^{2}}{%
2}x_{A}^{2}+\frac{1}{2}\left( \frac{dx_{B}}{dt}\right) ^{2}-\frac{\om_{B}^{2}}{%
2}y^{2}-\om_{AB}^{2}x_{A}x_{B}  \, . \label{toylagrangian}
\end{equation}%
In agreement with the previous discussion, we regard the term $L%
_{int}=\om_{AB}^{2}x_{A}x_{B}$ as an interaction, where $\om_{AB}^{2}$ (with
dimension Energy$^{2}$) plays the role of the coupling constant. Hence, the
fields in the interaction picture take the form: 
\begin{eqnarray}
x_{A}(t) &=&\frac{1}{\sqrt{2\om_{A}}}\left( a_{A}e^{-i\om_{A}t}+a_{A}^{\dagger
}e^{i\om_{A}t}\right) \text{ ,}  \label{xex} \\[2mm]
x_{B}(t) &=&\frac{1}{\sqrt{2\om_{B}}}\left( a_{B}e^{-i\om_{B}t}+a_{B}^{\dagger
}e^{i\om_{B}t}\right) \text{ ,}  \label{yex}
\end{eqnarray}%
in which the creation and annihilation operators (with usual commutation
relations $[a_{A},a_{A}^{\dagger }]=[a_{B},a_{B}^{\dagger }]=1$ and zero
otherwise) have been introduced.

We can safely perform calculations by means of the formula \eqref{dyfor2}, taking $H_{int}(t)=\om_{AB}^{2}x_{A}(t)x_{B}(t)$. 
As initial state $t_{i}$, we consider an excitation along the $A$-direction: 
$\vert A\rangle =a_{A}^{\dagger }\vert 0\rangle $.$\ $%
We then evaluate the probability that the state has changed at the time $%
t_{f}>t_{i},$ a situation that roughly speaking corresponds to a decay of
the initial state. The first possible transition is the mixing $\vert
A\rangle =a_{A}^{\dagger }\vert 0\rangle \rightarrow
a_{B}^{\dagger }\vert 0\rangle =\vert B\rangle $
driven by the interaction term. The corresponding amplitude reads: 
\begin{eqnarray}
\langle B\vert U(t_{f},t_{i})\vert A\rangle 
&=&\langle 0\vert a_{B}U(t_{f},t_{i})a_{A}^{\dagger }\vert
0\rangle =-i\frac{\om_{AB}^{2}}{\sqrt{2\om_{A}}\sqrt{2\om_{B}}}%
\int_{t_{i}}^{t_{f}}dt_{1}e^{-i(\om_{A}-\om_{B})t_{1}}  \notag \\
&=&\frac{\om_{AB}^{2}}{\sqrt{2\om_{A}}\sqrt{2\om_{B}}}\frac{%
e^{-i(\om_{A}-\om_{B})t_{f}}-e^{-i(\om_{A}-\om_{B})t_{i}}}{(\om_{A}-\om_{B})}\text{ .}
\end{eqnarray}%
Hence, the probability for this \textquotedblleft
transition\textquotedblright\ to happen is:%
\begin{equation}
\mathcal{P}_{A\rightarrow B}(\De t)=\frac{\om_{AB}^{4}}{\om_{A}\om_{B}}\frac{\sin^{2}\left[ 
\frac{(\om_{A}-\om_{B})\Delta t}{2}\right] }{(\om_{A}-\om_{B})^{2}}\text{ } \, , \qquad
\Delta t=t_{f}-t_{i} \, .
\end{equation}%
The formula includes an oscillation whose frequency is proportional to the
frequency differences, that we shall call the ``low frequency'' term. Note, for
short times $\mathcal{P}_{A\rightarrow B}(\De t)\simeq \frac{\om_{AB}^{4}\Delta t^{2}}{%
4\om_{A}\om_{B}}.$

There is, however, at first order another possible transition: $%
a_{A}^{\dagger }\vert 0\rangle \rightarrow \frac{\left(
a_{A}^{\dagger }\right) ^{2}}{\sqrt{2}}a_{B}^{\dagger }\left\vert
0\right\rangle ,$ that is a single excitation along $A$ converts into $AAB.$
The corresponding amplitude reads:%
\begin{eqnarray}
\frac{1}{\sqrt{2}}\langle 0\vert
a_{B}a_{A}^{2} \, U(t_{f},t_{i})\, a_{A}^{\dagger }\vert 0\rangle  &=&-i%
\frac{\sqrt{2}\om_{AB}^{2}}{\sqrt{2\om_{A}}\sqrt{2\om_{B}}}%
\int_{t_{i}}^{t_{f}}dt_{1}e^{-i(\om_{A}+\om_{B})t}  \notag \\
&=&\frac{\sqrt{2}\om_{AB}^{2}}{\sqrt{2\om_{1}}\sqrt{2\om_{2}}}\frac{%
e^{-i(\om_{A}+\om_{B})t_{f}}-e^{-i(\om_{A}+\om_{B})t_{i}}}{(\om_{A}+\om_{B})}\text{ ,}
\end{eqnarray}%
hence%
\begin{equation}
\mathcal{P}_{A\rightarrow AAB}(\De t)=\frac{2\om_{AB}^{4}}{\om_{A}\om_{B}}\frac{\sin ^{2}\left[ 
\frac{(\om_{A}+\om_{B})\Delta t}{2}\right] }{(\om_{A}+\om_{B})^{2}}\text{ ,}
\end{equation}%
which involves the sum of the frequencies and is denoted as the `high
frequency' term. For short times, $P^{A\rightarrow AAB}(\De t)\simeq \frac{%
\om_{AB}^{4}t^{2}}{2\om_{A}\om_{B}}$.

Summarizing, the total transition probability (in other words, the
 $A$ transition probability (loosely
speaking its decay probability) is given as the sum of both terms.%
\begin{equation}
\mathcal{P}_{D}^{A}(\De t) \ =\ \mathcal{P}_{A\rightarrow B}(\De t)+\mathcal{P}_{A\rightarrow AAB}(\De t)\ =\ \frac{\om_{AB}^{4}}{%
\om_{A}\om_{B}}\left[ \frac{\sin ^{2}\left[ \frac{(\om_{A}-\om_{B})\Delta t}{2}%
\right] }{(\om_{A}-\om_{B})^{2}}+2\frac{\sin ^{2}\left[ \frac{%
(\om_{A}+\om_{B})\Delta t}{2}\right] }{(\om_{A}+\om_{B})^{2}}\right] \, .
\end{equation}%
For short times, $\mathcal{P}_{D}^{A}(\De t)\simeq \frac{3\om_{AB}^{4}\Delta
t^{2}}{4\om_{A}\om_{B}}.$

Similarly, one can easily calculate within the same framework the survival
probability. To this end we need to evaluate 
\begin{equation}
\langle A\vert U(t_{f},t_{i})\vert A\rangle
=\langle 0\vert a_{A}U(t_{f},t_{i})a_{A}^{\dagger }\vert
0\rangle \, .
\end{equation}%
Up to the second order we get:

\begin{equation}
\langle A\vert U(t_{f},t_{i})\vert A\rangle =1-i \, %
\mathcal{T}\langle 0\vert
a_{A}\int_{t_{i}}^{t_{f}}dt_{1}H_{int}(t_{1})%
\int_{t_{i}}^{t_{1}}dt_{2}H_{int}(t_{2})a_{A}^{\dagger }\vert
0\rangle \, .
\end{equation}%
Upon using the equality%
\begin{equation}
\langle 0\vert a_{A}H_{int}(t_{1})H_{int}(t_{2})a_{A}^{\dagger
}\vert 0\rangle =\frac{2\om_{AB}^{4}}{4\om_{A}\om_{B}}%
e^{-i(\om_{A}+\om_{B})t_{1}}e^{i(\om_{A}+\om_{B})t_{2}}+\frac{\om_{AB}^{4}}{4\om_{A}\om_{B}%
}e^{i(\om_{A}-\om_{B})t_{1}}e^{-i(\om_{A}-\om_{B})t_{2}} \, , 
\end{equation}%
one gets the survival probability of the state $\vert A\rangle $
as:

\begin{align}
\mathcal{P}_{A \rightarrow A}(\De t)& =\left\vert 1-\frac{\om_{AB}^{4}}{4\om_{A}\om_{B}}\left[ 2\frac{t}{%
i(\om_{A}+\om_{B})}-2\frac{e^{-i(\om_{A}+\om_{B})\Delta t}-1}{(\om_{A}+\om_{B})^{2}}+%
\frac{t}{-i(\om_{A}-\om_{B})}-\frac{e^{i(\om_{A}-\om_{B})\Delta t}-1}{%
(\om_{A}-\om_{B})^{2}}\right] \right\vert ^{2} \\
& =\left\vert 1-R-iI\right\vert
^{2}=(1-R-iI)(1-R+iI)=1-R+iI-R+R^{2}-iRI-iI-iIR+I^{2} \\
& =1-2R+... \ ,
\end{align}%
where $R$ and $I$ are real. In particular:

\begin{eqnarray}
R &=&\frac{\om_{AB}^{4}}{4\om_{A}\om_{B}}\left( \frac{1-\cos \left[
(\om_{A}-\om_{B})\Delta t\right] }{(\om_{A}-\om_{B})^{2}}+2\frac{1-\cos \left[
(\om_{A}+\om_{B})\Delta t\right] }{(\om_{A}+\om_{B})^{2}}\right)   \notag \\
&=&\frac{\om_{AB}^{4}}{2\om_{A}\om_{B}}\left( \frac{\sin ^{2}\left[ \frac{%
(\om_{A}-\om_{B})\Delta t}{2}\right] }{(\om_{A}-\om_{B})^{2}}+2\frac{\sin ^{2}\left[ 
\frac{(\om_{A}+\om_{B})\Delta t}{2}\right] }{(\om_{A}+\om_{B})^{2}}\right) \, ,
\end{eqnarray}%
hence%
\begin{equation}
\mathcal{P}_{S}^{A}(\De t) = \mathcal{P}_{A \rightarrow A}(\De t) \ = \ 1-\frac{\om_{AB}^{4}}{\om_{A}\om_{B}}\left( \frac{\sin ^{2}\left[ 
\frac{(\om_{A}-\om_{B})\Delta t}{2}\right] }{(\om_{A}-\om_{B})^{2}}+2\frac{\sin ^{2}%
\left[ \frac{(\om_{A}+\om_{B})\Delta t}{2}\right] }{(\om_{A}+\om_{B})^{2}}\right) \, ,  
\end{equation}%
which leads to (at order $g^{2}$):%
\begin{equation}
\mathcal{P}_{S}^{A}(\De t)+\mathcal{P}_{D}^{A}(\De t)\ =\ 1 \, ,
\end{equation}%
for each $t,$ as it must. For small $t,$ $p_{S}^{A}(t)\simeq 1-\frac{%
\om_{AB}^{4}}{\om_{A}\om_{B}}\left( 2\frac{\Delta t^{2}}{4}+\frac{\Delta t^{2}}{4}%
\right) =1-\frac{3\om_{AB}^{4}\Delta t^{2}}{4\om_{A}\om_{B}}.$ 

Of course, the present problem can be also solved by introducing the
rotation 
\begin{equation}
\left( 
\begin{array}{c}
x_{1} \\ 
x_{2}%
\end{array}%
\right) =\left( 
\begin{array}{cc}
\cos \theta  & -\sin \theta  \\ 
\sin \theta  & \cos \theta 
\end{array}%
\right) \left( 
\begin{array}{c}
x_{A} \\ 
x_{B}%
\end{array}%
\right) 
\end{equation}%
with%
\begin{equation}
\theta =\frac{1}{2}\arctan \frac{2\om_{AB}^{2}}{\om_{B}^{2}-\om_{A}^{2}} \, ,
\end{equation}%
and%
\begin{align}
\om_{1}^{2}& =\om_{A}^{2}\cos ^{2}\theta +\om_{B}^{2}\sin ^{2}\theta
-\om_{AB}^{2}\sin(2\theta )\text{ }, \\
\om_{2}^{2}& =\om_{A}^{2}\sin ^{2}\theta +\om_{B}^{2}\cos ^{2}\theta
+\om_{AB}^{2}\sin (2\theta )\text{ }\,, \\
\om_{A}^{2}& =\om_{1}^{2}\cos ^{2}\theta +\om_{2}^{2}\sin ^{2}\theta \text{ }\,, \\
\om_{B}^{2}& =\om_{1}^{2}\sin ^{2}\theta +\om_{2}^{2}\cos ^{2}\theta \text{ }\,%
\text{.}
\end{align}%
The position operators become 
\begin{eqnarray}
x_{1}(t) &=&\frac{1}{\sqrt{2\om_{1}}}\left( a_{1}e^{-i\om_{1}t}+a_{1}^{\dagger
}e^{i\om_{1}t}\right) \, ,  \\[2mm]
x_{2}(t) &=&\frac{1}{\sqrt{2\om_{2}}}\left( a_{2}e^{-i\om_{2}t}+a_{2}^{\dagger
}e^{i\om_{2}t}\right)  \, .
\end{eqnarray}%
Upon denoting $\left\vert \Omega \right\rangle $ as the vacuum of the full
Hamiltonian ($a_{1}\left\vert \Omega \right\rangle =a_{2}\left\vert \Omega
\right\rangle =0$), one may also consider the state 
\begin{equation}
\left\vert a\right\rangle =\cos \theta a_{1}^{\dagger }\left\vert \Omega
\right\rangle +\sin \theta a_{2}^{\dagger }\left\vert \Omega \right\rangle \, , 
\end{equation}%
yet it is clear that $\left\vert a\right\rangle \neq \left\vert
A\right\rangle =a_{A}^{\dagger }\left\vert a\right\rangle $,

In terms of $\left\vert a\right\rangle $, the survival probability takes the
form: 
\begin{equation}
\mathcal{P}_{S}^{a}(\De t) \ = \ 1-\sin ^{2}2\theta \sin ^{2}\left[ \frac{(\om_{1}-\om_{2})\Delta t}{%
2}\right]  \, \text{.}
\end{equation}%
In the limit of small $\theta ,$ the previous expression is approximated by:%
\begin{equation}
\mathcal{P}_{S}^{a}(\De t)  \ \simeq \ 1-\frac{4\om_{AB}^{4}}{\left( \om_{B}^{2}-\om_{A}^{2}\right) ^{2}}%
\sin ^{2}\left[ \frac{(\om_{A}-\om_{B})\Delta t}{2}\right] =1-\frac{4\om_{AB}^{4}}{%
\left( \om_{B}+\om_{A}\right) ^{2}}\frac{\sin ^{2}\left[ \frac{%
(\om_{A}-\om_{B})\Delta t}{2}\right] }{\left( \om_{B}-\om_{A}\right) ^{2}} \, \text{ .}
\end{equation}%
We then realize that the functions $\mathcal{P}_{S}^{a}(\De t) $ and $\mathcal{P}_{S}^{A}(\De t) $ are
different in various ways. First, the expression $\mathcal{P}_{S}^{a}(\De t) $ contains
only the low frequency term but not the high frequency one. Second, the
ratio of the coefficients in front of the terms with frequency $\left(
\om_{A}-\om_{B}\right) $ reads
$4\om_{A}\om_{B}/\left( \om_{A}+\om_{B}\right)^{2}$, 
which is in general different from unity (it approaches for it in the limit
of equal bare masses). This discrepancy is due to the fact that the states $%
\left\vert a\right\rangle $ and $\left\vert A\right\rangle $ are different,
thus they have different survival probabilities. In the framework of QM, one
may ``engineer'' both initial states. In QFT it is different, and it is not a
priori clear to what the field $x_{A}$ corresponds to. 

%%%%%%%%%%%%%%%%%%%%%%%%%%%%%%%%%%%%%%%%%%%%%%%%%%%%%%%%%%%%%%%%%%%%%%%%%%%%%%%%%%%%
\section{Scalar field mixing in the interaction picture} \label{bosesec}

We now move from QM to QFT. To this end, we investigate the mixing for
scalar fields in the interaction picture. 

Let us  consider two fields $\phi_{A}=\phi_{A}(t,\mathbf{x})$ and $\phi_{B}=\phi_{B}(t,%
\mathbf{x})$ that correspond to our flavor bare states $A$ and $B,$  whose
Lagrangian density is given by
\begin{equation}
\mathcal{L}=\frac{1}{2}\left( \partial _{\alpha }\phi_{A}\right) ^{2}-\frac{%
m_{A}^{2}}{2}\phi_{A}^{2}+\frac{1}{2}\left( \partial _{\alpha }\phi_{B}\right)
^{2}-\frac{m_{B}^{2}}{2}\phi_{B}^{2}-m_{AB}^{2}\phi_{A}\phi_{B}\text{.}
\end{equation}%
The Hamiltonian density is
\begin{equation}
\mathcal{H}=\frac{\pi _{A}}{2}+\frac{\pi _{B}}{2}+\frac{1}{2}\left(
\nabla \phi_{A}\right) ^{2}+\frac{1}{2}\left( \nabla
\phi_{B}\right) ^{2}+\frac{m_{A}^{2}}{2}\phi_{A}^{2}+\frac{m_{A}^{2}}{2}%
\phi_{B}^{2}+m_{AB}^{2}\phi_{A}\phi_{B} \, ,
\end{equation}%
with $\pi _{A}=\partial _{t}\phi_{A}$ and $\pi _{B}=\partial _{t}\phi_{B}.$ We
regard the mixing term as a perturbation, thus:

\begin{equation}
\mathcal{H}_{0}=\frac{\pi _{A}}{2}+\frac{\pi _{B}}{2}+\frac{1}{2}\left(
\nabla \phi_{A}\right) ^{2}+\frac{1}{2}\left( \nabla
\phi_{B}\right) ^{2}+\frac{m_{A}^{2}}{2}\phi_{A}^{2}+\frac{m_{A}^{2}}{2}\phi_{B}^{2} \, 
 , \qquad \mathcal{H}_{int}=m_{AB}^{2}\phi_{A}\phi_{B} \, \text{.}
\end{equation}%
Upon quantizing the system, in the interaction picture we get for the field $%
A$:
\begin{align}
\phi_{A}(x)& =\phi_{A}(t,\mathbf{x})=\frac{1}{\sqrt{V}}\sum_{\textbf{k}=2\pi 
\mathbf{n}/L}\frac{1}{\sqrt{2\omega _{\G k,A}}}\left( a_{\G k,A%
}e^{-ikx}+a_{\G k,A}^{\dagger }e^{ikx}\right)  \, ,\\
\pi _{A}(x)& =\pi _{A}(t,\mathbf{x})=\frac{-i}{\sqrt{V}}\sum_{\textbf{k}%
=2\pi \mathbf{n}/L}\sqrt{\frac{\omega_{\G k,A}}{2}}\left( a_{\G k,%
A}e^{-ikx}-a_{\G k,A}^{\dagger }e^{ikx}\right) \, ,
\end{align}%
with $k^{0}=\omega_{\G k,A}=\sqrt{\textbf{k}^{2}+m_{A}^{2}}.$ The
commutation relation $\left[ \phi_{A}(t,\mathbf{x}),\pi _{A}(t,\mathbf{y})%
\right] =\frac{1}{V}\sum_{\textbf{k}}e^{i\mathbf{k\cdot (x-y)}}=i\delta_V(%
\mathbf{x-y})$ implies that $\left[ a_{\G k,A},a_{\G p,A}%
^{\dagger }\right] =\delta_{\G k,\G p}$, zero
otherwise. Analogous expressions hold for $\phi_{B}(x)$ and $\pi _{B}(x).$ 

The interacting Hamiltonian (in the interaction picture) reads:

\begin{align} \non 
H_{int}(t)& =\intx_{1}\mathcal{H}_{int}(x)=\sum_{\textbf{q}}\frac{%
m_{AB}^{2}}{\sqrt{2\omega_{\G q,A}}\sqrt{2\omega_{\G q,B}}}\left( a_{\G q,A} a_{\G q,B}^{\dagger }e^{-i\left( \omega_{\G q,A} 
 -\omega_{\G q,B}\ri)t} \right.
 \\
& \left. + a_{\G q,A}^{\dagger
}a_{\G q,B} e^{i\left( \omega_{\G q,A} -\omega_{\G q,B}%
\ri)t}  + a_{\G q,A}a_{-\G q,B}e^{-i\left( \omega_{\G q,A}%
 +\omega_{\G q,B}\ri)t} + a_{\G q,A}^{\dagger}a_{-\G q,B}%
^{\dagger }e^{i\left( \omega_{\G q,A} +\omega_{\G q,B}%
\ri)t}\right) \, .
\end{align}

Next, we define the ``flavor state'' $A$ with three-momentum $\textbf{p}$ as: 
\begin{equation}
\left\vert A,\textbf{p}\right\rangle =a_{\G p,A}^{\dagger }\left\vert
0\right\rangle \text{ .}
\end{equation}%
Assuming that such a state is created at $t=0$, we evaluate the probability
that it has transformed into a different state at the time $t>0$ or,
conversely, that it has not changed.

For the case of the transition into a different state, let us first
calculate the probability amplitude for the transition $\left\vert A,\mathbf{%
p}\right\rangle \rightarrow \left\vert B,\textbf{k}\right\rangle $:

\begin{align}\non
\mathcal{A}_{A\rightarrow B}\left( \textbf{p},\G{k}; t_i,t_f\right) &
=\langle B,\textbf{k}\vert U(t_{f},t_{i})\vert A,\textbf{p}%
\rangle =-i\int_{t_{i}}^{t_{f}}dt_{1}\langle 0\vert a_{\G k,B}%
\,H_{int}(t_{1})\,a_{\G p,A}^{\dagger }\vert 0\rangle
+... \\
& =\frac{m_{AB}^{2}}{\sqrt{2\omega_{\G p,A}}\sqrt{2\omega_{\G k,B}%
}}\delta _{\textbf{k},\textbf{p}}\frac{e^{-i\left( \omega_{\G p,A}%
 -\omega_{\G k,B}\ri)t_{f}}-e^{-i\left( \omega_{\G p,A}%
 -\omega_{\G k,B}\ri)t_{i}}}{ \om_{\G p,A}%
-\omega_{\G k,B}}\, .
\end{align}%
The probability that a particle $A$ with momentum $\textbf{p}$ converts into
a particle $B$ is obtained upon summing over the density of final states $%
\sum_{\textbf{k}}$ :%
\begin{equation} \label{probsum}
\mathcal{P}_{A\rightarrow B}(\G{p}; \De t)=\sum_{\textbf{k}}\vert
\mathcal{A}_{A\rightarrow B}\left( \textbf{p},\G{k};t_i,t_f\right) \vert ^{2}=%
\frac{m_{AB}^{4}}{\omega_{\G p,A}\omega_{\G p,B}}\frac{\sin
^{2}\left[ \frac{\left( \omega_{\G p,A} -\omega_{\G p,B}%
\ri)\Delta t}{2}\right] }{\left( \omega_{\G p,A} -\omega_{\G p,B}%
\ri)^{2}}\, .
\end{equation}%
The result is finite and well behaved. Note, for short time $\mathcal{P}_{A\rightarrow
B}(\G{p}; \De t)\simeq \frac{m_{AB}^{4}}{4\omega_{\G p,A}%
\omega_{\G p,B}}t^{2}.$

Yet, other transitions are possible. Namely, we may have the transition $%
A\rightarrow AAB$ of the type%
\begin{equation}
\left\vert A,\textbf{p}\right\rangle \rightarrow \left\vert A,\textbf{k}%
_{1}\right\rangle \left\vert A,\textbf{k}_{2}\right\rangle \left\vert B,%
\textbf{k}_{3}\right\rangle =a_{A,\textbf{k}_{1}}^{\dagger }a_{A,\textbf{k}%
_{2}}^{\dag }a_{B,\textbf{k}_{3}}^{\dag }\left\vert 0\right\rangle \text{ ,}
\end{equation}%
where the two emitted $A$ particles have different momentum, $\textbf{k}%
_{1}\neq \textbf{k}_{2}.$ The corresponding amplitude of this process reads%
\begin{equation}
\mathcal{A}_{A\rightarrow AAB}^{\textbf{k}_{1}\neq \textbf{k}_{2}}(\textbf{p}%
,\textbf{k}_{1},\textbf{k}_{2},\textbf{k}_{3};t_i,t_f)=-i\int_{t_{i}}^{t_{f}}dt_{1}\langle 0\vert a_{A,\textbf{k}%
_{1}}a_{A,\textbf{k}_{2}}a_{B,\textbf{k}_{3}}H_{int}(t_{1})a_{A,\textbf{p}%
}^{\dagger }\vert 0\rangle \text{ .}
\end{equation}%
After an explicit calculation up to first order, its squared modulus turns
out to be: 
\begin{equation}
\vert \mathcal{A}_{A\rightarrow AAB}^{\textbf{k}_{1}\neq \textbf{k}%
_{2}}(\textbf{p},\textbf{k}_{1},\textbf{k}_{2},\textbf{k}%
_{3},t_i,t_f)\vert ^{2}=\frac{m_{AB}^{4}}{\omega_{\G k_3,A}%
\omega _{\G k_3,B}}\frac{\sin^{2}\Big[ \frac{\left( \omega
_{\textbf{k}_{3},A} +\omega _{\G k_3,B}\ri)\Delta t}{2}\Big] }{%
\left( \omega_{\textbf{k}_{3},A} +\omega _{\G k_3,B}\ri)^{2}}%
\left( \delta _{\textbf{k}_{1},\textbf{p}}\delta _{\textbf{k}_{2},-\textbf{k}%
_{3}}+\delta _{\textbf{k}_{1},-\textbf{k}_{3}}\delta _{\textbf{k}_{2},%
\textbf{p}}\right) \text{ .}
\end{equation}%
Next, one needs to sum over final states $\textbf{k}_{1},\textbf{k}_{2},%
\textbf{k}_{3}.$ leading to the probability:%
\begin{eqnarray}\non
\mathcal{P}^{\G k_1 \neq \G k_2}_{A\rightarrow AAB}(\G{p};\De t) &=&\text{ }\frac{1}{2}\sum_{%
\textbf{k}_{1},\textbf{k}_{2},\textbf{k}_{3}}\left\vert 
\mathcal{A}_{A\rightarrow AAB}^{\textbf{k}_{1}\neq \textbf{k}_{2}}(\textbf{p}%
,\textbf{k}_{1},\textbf{k}_{2},\textbf{k}_{3},t_i,t_f)\right\vert ^{2} \\
&=&\sum_{\textbf{k}_{3}}\frac{m_{AB}^{4}}{\omega _{\G k_3,A}\omega
_{\G k_3,B}}\frac{\sin^{2}\Big[ \frac{\left( \omega _{\G k_3,A}%
 +\omega _{\G k_3,B}\ri)\Delta t}{2}\Big] }{\left( \omega
_{\textbf{k}_{3},A} +\omega _{\G k_3,B}\ri)^{2}}-\frac{%
m_{AB}^{4}}{\omega _{\G p,A}\omega _{\G p,B}}\frac{\sin ^{2}%
\left[ \frac{\left( \omega _{\G p,A} +\omega _{\G p,B}\ri)%
\Delta t}{2}\right] }{\left( \omega _{\G p,A} +\omega _{\G p,B}%
\ri)^{2}} \, ,
\end{eqnarray}%
where the factor $1/2$ in front of the sum takes into account that the two $A
$ in the final state are identical bosons. The subtracted term in the last
equation is due to the condition $\textbf{k}_{2}\neq \textbf{k}_{1}.$ Note,
the sum term diverges, thus a certain cutoff is implicitly introduced so to
keep the intermediate results finite (which is then sent to infinity at the
very end of the calculation). When the volume is sufficiently large, the
previous expression becomes%
\begin{equation}
\mathcal{P}^{\G k_1 \neq \G k_2}_{A\rightarrow AAB}(\G p;\Delta t) =V\int \!\! \frac{\dr^{3}\G k_{3}}{(2\pi
)^{3}} \, \frac{m_{AB}^{4}}{\omega _{\G k_3,A}\omega
_{\G k_3,B}}\frac{\sin^{2}\Big[ \frac{\left( \omega _{\G k_3,A}%
 +\omega _{\G k_3,B}\ri)\Delta t}{2}\Big] }{\left( \omega
_{\textbf{k}_{3},A} +\omega _{\G k_3,B}\ri)^{2}}-\frac{%
m_{AB}^{4}}{\omega _{\G p,A}\omega _{\G p,B}}\frac{\sin ^{2}%
\left[ \frac{\left( \omega_{\G p,A} +\omega _{\G p,B}\ri)%
\Delta t}{2}\right] }{\left( \omega _{\G p,A} +\omega _{\G p,B}%
\ri)^{2}} \, , 
\label{AABdiff}
\end{equation}%
where, again, a cutoff is implicit in the integral over $k_{3}$. The term
proportional to $V$ is a typical vacuum term that needs to be subtracted.
However, the second term in Eq.(\ref{AABdiff}) needs to be kept, see below.

The last possible transition is the case in which $\textbf{k}_{2}= 
\textbf{k}_{1},$ thus%
\begin{equation}
\vert A,\textbf{p}\rangle \rightarrow \frac{1}{\sqrt{2}}a_{\G k_1,%
A}^{\dagger }a_{\textbf{k}_{1},A}^{\dag }a_{\textbf{k}_{3},B%
}^{\dag }\vert 0\rangle \text{ .}
\end{equation}%
The amplitude at first order is 
\begin{equation}
\mathcal{A}_{A\rightarrow AAB}^{\textbf{k}_{1}=\textbf{k}_{2}}(\textbf{p},%
\textbf{k}_{1},\textbf{k}_{3};t_i,t_f)=-\frac{i}{\sqrt{2}}%
\int_{t_{i}}^{t_{f}}dt_{1}\langle 0\vert a_{\textbf{k}_{1},A}a_{%
\textbf{k}_{1},A}a_{\textbf{k}_{3},B}H_{int}(t_{1})a_{\textbf{p},A}^{\dagger
}\vert 0\rangle \, , 
\end{equation}%
whose squared modulus is%
\begin{equation}
\vert \mathcal{A}_{A\rightarrow AAB}^{\textbf{k}_{1}=\textbf{k}_{2}}(%
\textbf{p},\textbf{k}_{1},\textbf{k}_{3};t_i,t_f)\vert ^{2}=\delta _{%
\textbf{k}_{1},\textbf{p}}\delta _{\textbf{k}_{1},\mathbf{-k}_{3}}2\frac{%
m_{AB}^{4}}{\omega _{\G p,A}\omega _{\G p,B}}\frac{\sin ^{2}%
\left[ \frac{\left( \omega _{\G p,A} +\omega _{\G p,B}\ri)%
\Delta t}{2}\right] }{\left( \omega _{\G p,A} +\omega _{\G p,B}%
\ri)^{2}} \, .
\end{equation}%
Upon summing over the final momenta $\textbf{k}_{1},\textbf{k}_{3}$:%
\begin{equation}
\mathcal{P}^{\G k_1 = \G k_2}_{A\rightarrow AAB}(\G{p},\De t) =\text{ }\sum_{\textbf{k}_{1},%
\textbf{k}_{3}}\left\vert \mathcal{A}_{A\rightarrow AAB}^{\textbf{k}_{1}=%
\textbf{k}_{2}}(\textbf{p},\textbf{k}_{1},\textbf{k}_{3};t_i,t_f)\right\vert ^{2}=2\frac{%
m_{AB}^{4}}{\omega _{\G p,A}\omega _{\G p,B}}\frac{\sin ^{2}%
\left[ \frac{\left( \omega _{\G p,A} +\omega _{\G p,B}\ri)%
\Delta t}{2}\right] }{\left( \omega _{\G p,A} +\omega _{\G p,B}%
\ri)^{2}}\text{ .}
\end{equation}%
Note, the factor $2$ appears just as in the QM toy model of Section\ref{childsec}. 

Putting all the pieces together, the total probability of $A$ going
into something else (thus, a decay probability) is:%
\begin{eqnarray}\non
\mathcal{P}_{D}^{A}(\G{p}; \De t) &=&\mathcal{P}_{A\rightarrow B}(\G{p}; \De t) +\mathcal{P}^{\G k_1 \neq \G k_2}_{A\rightarrow AAB}(\G{p}; \De t) +\mathcal{P}^{\G k_1 = \G k_2}_{A\rightarrow AAB}(\G{p}; \De t)  
\\ [2mm] \non
&=&\frac{%
m_{AB}^{4}}{\omega _{\G p,A}\omega _{\G p,B}}\frac{\sin ^{2}%
\left[ \frac{\left( \omega _{\G p,A} -\omega _{\G p,B}\ri)%
\Delta t}{2}\right] }{\left( \omega _{\G p,A} -\omega _{\G p,B}%
\ri)^{2}}+\frac{%
m_{AB}^{4}}{\omega _{\G p,A}\omega _{\G p,B}}\frac{\sin ^{2}%
\left[ \frac{\left( \omega _{\G p,A} +\omega _{\G p,B}\ri)%
\Delta t}{2}\right] }{\left( \omega _{\G p,A} +\omega _{\G p,B}%
\ri)^{2}}\\
&&+V\int \!\! \frac{\dr^{3}\G k_{3}}{(2\pi )^{3}}\frac{m_{AB}^{4}}{\omega _{\G k_3,A}%
\omega _{\G k_3,B}}\frac{\sin ^{2}\Big[ \frac{\left( \omega
_{\G k_3,A}+\omega _{\G k_3,B}\ri)\Delta t}{2}\Big] }{%
\left( \omega _{\G k_3, A}+\omega _{\G k_3,B}\ri)^{2}}%
\text{ .}
\end{eqnarray}%
The factor $2$ of $\mathcal{P}^{\G k_1 = \G k_2}_{A\rightarrow AAB}(\G{p}; \De t)$ combines
with the factor $-1$ in $\mathcal{P}^{\G k_1 \neq \G k_2}_{A\rightarrow AAB}(\G{p}; \De t)$ in
order to give the same factor in front of the high-frequency term. Finally,
the term proportional to $V$, being a vacuum term, is subtracted. Then the
probability that the oscillation takes place up to second order is: 
\begin{equation}
\mathcal{P}_{D}^{A}(\G{p};\Delta t) =\frac{m_{AB}^{4}}{\omega _{\G p,A}%
\omega _{\G p,B}}\left( \frac{\sin ^{2}%
\left[ \frac{\left( \omega _{\G p,A} -\omega _{\G p,B}\ri)%
\Delta t}{2}\right] }{\left( \omega _{\G p,A} -\omega _{\G p,B}%
\ri)^{2}}+\frac{\sin ^{2}%
\left[ \frac{\left( \omega _{\G p,A} +\omega _{\G p,B}\ri)%
\Delta t}{2}\right] }{\left( \omega _{\G p,A} +\omega _{\G p,B}%
\ri)^{2}}\right) \text{ .}
\end{equation}

It is important to verify the correctness of the previous result. Just as in
the QM toy model, one needs to check the survival probability of the state $%
\left\vert A,\textbf{p}\right\rangle $. To this end, we calculate the
probability of the transition%
\begin{equation}
\left\vert A,\textbf{p}\right\rangle \rightarrow \left\vert A,\textbf{k}%
\right\rangle \, ,
\end{equation}%
and then sum over $\mathbf{k.}$ The corresponding amplitude, up to second
order, reads: 
\begin{equation}
\mathcal{A}_{A\rightarrow A}(\textbf{p},\textbf{k};t_i,t_f)=\langle
0\vert a_{\textbf{k},A}^{\dagger }U(t_{f},t_{i})a_{\textbf{p},A%
}^{\dagger }\vert 0\rangle =\delta _{\mathbf{k,p}%
}+(-i)^{2}\int_{t_{i}}^{t_{f}}dt_{1}\int_{t_{i}}^{t_{1}}dt_{2}\langle
0\vert a_{\textbf{k},A}H_{int}(t_{1})H_{int}(t_{2})a_{\textbf{p},A%
}^{\dagger }\vert 0\rangle \text{.}  
\end{equation}%
Its modulus square takes the form:%
\begin{eqnarray}
\left\vert \mathcal{A}_{A\rightarrow A}(\textbf{p},\textbf{k};t_i,t_f)\right\vert ^{2}  
&=&\delta _{\mathbf{pk}}\left( 1-\frac{%
m_{AB}^{4}}{\omega _{\G p,A}\omega _{\G p,B}}\frac{\sin ^{2}%
\left[ \frac{\left( \omega _{\G p,A} -\omega _{\G p,B}\ri)%
\Delta t}{2}\right] }{\left( \omega _{\G p,A} -\omega _{\G p,B}%
\ri)^{2}}-\frac{m_{AB}^{4}%
}{\omega _{\G p,A}\omega _{\G p,B}}\frac{\sin ^{2}\left[ \frac{%
\left( \omega _{\G p,A} +\omega _{\G p,B}\ri)\Delta t}{2}%
\right] }{\left( \omega _{\G p,A} +\omega _{\G p,B}\ri)^{2}}%
\right)   \notag \\
&&-\delta _{\mathbf{pk}}\sum_{\textbf{q}_{1}}\frac{m_{AB}^{4}}{\omega _{\textbf{q}_{1},A}%
2\omega _{\textbf{q}_{1},B}}\frac{\sin ^{2}\Big[ \frac{%
\left( \omega _{\textbf{q}_{1},A} +\omega _{\textbf{q}_{1},B}\ri)\Delta t%
}{2}\Big] }{\left( \omega _{\textbf{q}_{1},A} +\omega _{\textbf{q}_{1},B}%
\ri)^{2}}+... \ , 
\end{eqnarray}%
where dots refer to higher order terms.

Then, upon summing over the final three-momentum $\textbf{k}$ and taking the
large volume limit, the survival probability reads (up to second order):%
\begin{eqnarray}
\mathcal{P}^A_{S}(\textbf{p};\Delta t) \ = \ \mathcal{P}_{A \rightarrow A}(\textbf{p;}\Delta t) &=&\sum_{\textbf{k}}\left\vert \mathcal{A}%
_{A\rightarrow A}(\textbf{p},\textbf{k};t_i,t_f)\right\vert ^{2}  \notag \\
&=&1-\frac{%
m_{AB}^{4}}{\omega _{\G p,A}\omega _{\G p,B}}\frac{\sin ^{2}%
\left[ \frac{\left( \omega _{\G p,A} -\omega _{\G p,B}\ri)%
\Delta t}{2}\right] }{\left( \omega _{\G p,A} -\omega _{\G p,B}%
\ri)^{2}}-\frac{m_{AB}^{4}%
}{\omega _{\G p,A}\omega _{\G p,B}}\frac{\sin ^{2}\left[ \frac{%
\left( \omega _{\G p,A} +\omega _{\G p,B}\ri)\Delta t}{2}%
\right] }{\left( \omega _{\G p,A} +\omega _{\G p,B}\ri)^{2}} \notag \\
&&-V\int \!\! \frac{\dr^{3}\G q_1}{(2\pi )^{3}} \, \frac{m_{AB}^{4}}{\omega _{\textbf{q}_{1},A}%
2\omega _{\textbf{q}_{1},B}}\frac{\sin^{2}\Big[ \frac{%
\left( \omega _{\textbf{q}_{1},A} +\omega _{\textbf{q}_{1},B}\ri)\Delta t%
}{2}\Big] }{\left( \omega _{\textbf{q}_{1},A} +\omega _{\textbf{q}_{1},B}%
\ri)^{2}} \, , 
\end{eqnarray}%
where the latter term corresponds, in diagrammatic term, to the disconnected
vacuum diagram with an $AB$ loop. This term coincides exactly with the one
obtained previously. Upon subtracting this term, we find:%
\begin{equation}
\mathcal{P}^A_{S}(\textbf{p};\Delta t)  \ = \ 1- \frac{m_{AB}^{4}}{\omega _{\G p,A}%
\omega _{\G p,B}}\left( \frac{\sin ^{2}%
\left[ \frac{\left( \omega _{\G p,A} -\omega _{\G p,B}\ri)%
\Delta t}{2}\right] }{\left( \omega _{\G p,A} -\omega _{\G p,B}%
\ri)^{2}}+\frac{\sin ^{2}%
\left[ \frac{\left( \omega _{\G p,A} +\omega _{\G p,B}\ri)%
\Delta t}{2}\right] }{\left( \omega _{\G p,A} +\omega _{\G p,B}%
\ri)^{2}}\right) \, , 
\label{pSAnorm}
\end{equation}%
with 
\begin{equation}
\mathcal{P}^A_{D}(\textbf{p};\Delta t) +\mathcal{P}^A_{S}(\textbf{p};\Delta t) \ =\ 1 \, , 
\end{equation}%
as it must. 

We thus obtain the probability that the flavor $A$ oscillates (or does not
oscillate) as the sum of two distinct term involving the low-frequency and
the high-frequency term, where the frequencies $\omega _{\G p,A}$ and 
$\omega _{\G p,B}$ depend on the chosen momentum $\mathbf{p.}$ Note,
the structure of the solution is very similar to the QM toy model besides
the factor $2$ of the high-frequency term. It turns out that only $1$
survives the process of renormalization.

Also in the scalar QFT case one may introduce a suitable diagonalization of
the fields%
\begin{equation}
\left( 
\begin{array}{c}
\phi_{1} \\ 
\phi_{2}%
\end{array}%
\right) =\left( 
\begin{array}{cc}
\cos \theta  & -\sin \theta  \\ 
\sin \theta  & \cos \theta 
\end{array}%
\right) \left( 
\begin{array}{c}
\phi_{A} \\ 
\phi_{B}%
\end{array}%
\right) \, , 
\end{equation}%
with%
\begin{equation}
\theta =\frac{1}{2}\arctan \frac{2m_{AB}^{2}}{m_{B}^{2}-m_{A}^{2}} \, .
\end{equation}%
The fields $\phi_{1}$ and $\phi_{2}$ contain the annihilation (creation) operators 
$a_{1,\mathbf{p,}}$ $a_{2,\mathbf{p,}}$ ($a_{1,\textbf{p}}^{\dagger }$,$a_{2,%
\textbf{p}}^{\dagger })$. If we introduce the ``Pontecorvo state'' 
\begin{equation}
\left\vert a,\textbf{p}\right\rangle =\cos \theta a_{1}^{\dagger }\left\vert
\Omega \right\rangle +\sin \theta a_{2}^{\dagger }\left\vert \Omega
\right\rangle \, , 
\end{equation}%
the corresponding survival probability takes the form: 
\begin{equation}
\mathcal{P}_S^a(\G p;\Delta t)\ = \ 1-\sin ^{2}2\theta \sin ^{2}\left[ \frac{(\omega
_{\G p,1}-\omega _{\G p,2})\Delta t}{2}\right] \, , 
\end{equation}%
with $\omega _{\G p,j}=\sqrt{|\textbf{p}|^{2}+m_{j}^{2}}$, $j=1,2$. In the limit of
small $\theta ,$ the previous expression is approximated by:%
\begin{equation}
\mathcal{P}_S^a(\G p;\Delta t) \simeq 1-\frac{4m_{AB}^{4}}{\left( m
_{B}^{2}-m_{\G A}^{2}\right) ^{2}}\sin ^{2}\left[ \frac{(\omega
_{\G p,A}-\omega _{\G p,B})\Delta t}{2}\right] \, ,
\end{equation}%
which, just as in the QM toy model, differs from Eq.(\ref{pSAnorm}) since
the high-frequency term is missing and because the factor in front of the
low-frequency one is not the same (but the two expressions degenerate for
large $\left\vert \textbf{p}\right\vert $)$.$ The difference is expected
because $\left\vert a,\textbf{p}\right\rangle \neq \left\vert A,\textbf{p}%
\right\rangle .$

%%%%%%%%%%%%%%%%%%%%%%%%%%%%%%%%%%%%%%%%%%%%%%%%%%%%%%%%%%%%%%%%%%%%%%%%%%%%%%%%%%%%%%%%%%%%%%%
\section{Neutrino oscillations in the interaction picture} \label{neutsec}
Let us now deal with the fermion case, which can be naturally applied to neutrino oscillations and it will be then referred as \emph{neutrino case}.
 
In the interaction picture $\nu_\si$ ($\si=e,\mu$), defined by the Lagrangian Eq.\eqref{neutr}, can be expanded as free fields, evolving under the action of $\mathcal{L}_0$:
\begin{eqnarray}
\nu_{\si}(x) = \frac{1}{\sqrt{V}} \sum_{\G k,r}\,  \left[ u_{{\bf k},\si}^{r}(t) \, \alpha_{{\bf k},\si}^{r} + v_{-{\bf k},\si}^{r}(t) \, \bt_{-{\bf k},\si}^{r\dag}   \right]  e^{i{\bf k}\cdot {\bf x}}  \, ,
\label{fieldex}
\end{eqnarray}
with $u^r_{{\bf k},\si}(t) \,= \, e^{- i \om_{\G k,\si} t}\, u^r_{{\bf k},\si}\;$,
$\;v^r_{{\bf k},\si}(t) \,= \, e^{ i \om_{\G k,\si} t}\, v^r_{{\bf k},\si}$,
 $\om_{\G k,\si}=\sqrt{|\G k|^2 + m_\si^2}$. Annihilation operators satisfy
\be \label{vacm}
\al^r_{\G k, \si}|0 \rangle = 0 = \beta _{{\bf k},\si}^{r} |0 \rangle \  .
\ee
The anticommutation relations are
\be  \label{CAR2} \{\al ^r_{{\bf k},\rho}, \al ^{s\dag }_{{\bf q},\si}\} = \de_{\G k \G q}\de _{rs}\de _{\rho \si}  \quad \, , \quad \{\bt^r_{{\bf k},\rho},
\bt^{s\dag }_{{\bf q},\si}\} =
\de_{\G k \G q} \de _{rs}\de _{\rho \si}, 
\ee
and the spinors are normalized so that
\bea \label{orth}
u^{r\dag}_{{\bf k},\rho} u^{s}_{{\bf k},\rho} =
v^{r\dag}_{{\bf k},\rho} v^{s}_{{\bf k},\rho} \ =  \ \de_{rs}
\quad, \quad u^{r\dag}_{{\bf k},\rho} v^{s}_{-{\bf k},\rho} = 0 \;.
\eea

As in the previous examples, the idea is to perform the perturbative calculation up the first in $m_{e \mu}$.
The interacting Hamiltonian reads: 

\bea
H_{int}(t)& = &  m_{e \mu}  \sum_{s,s'=1,2}\sum_{\G p}
 \Big[\bt^s_{\G p,\mu}\bt^{s\dag}_{\G p,e} \de_{s s'} W^*_\G p(t)+\al^{r\dag}_{\G p,\mu} \al^r_{\G p,e} \de_{s s'} W_\G p(t) \non \\[2mm]
& + & \bt^s_{-\G p,\mu}\al^{s'}_{e,\G p} \lf(Y^{s s'}_\G p(t)\ri)^*+\al^{s\dag}_{\G p,\mu}\bt^{s'\dag}_{-\G p,e} Y^{s s'}_\G p(t)\, + \, e \leftrightarrow \mu \Big] \, ,
\eea
where we defined
\bea
W_\G p(t) & = & \overline{u}^s_{\G p,\mu} u^s_{\G p,e} e^{i\left( \omega_{\G k,\mu}-\omega_{\G k,e}\ri)t} \ = \ W_\G p \,  e^{i\left( \omega_{\G p,\mu}-\omega_{\G p,e}\ri)t}  \\[2mm]
Y^{s s'}_\G p(t) & = &  \, \overline{u}^{s}_{\G p,\mu} v^{s'}_{-\G p,e} e^{i\left( \omega_{\G k,\mu}+\omega_{\G k,e}\ri)t} \ = \ Y^{s s'}_\G p e^{i\left( \omega_{\G p,\mu}+\omega_{\G p,e}\ri)t}
\eea
Explicitly
\bea
W_\G p & = & \sqrt{\frac{\lf(\omega_{\G p,e}+m_{e}\ri)\lf(\omega_{\G p,\mu}+m_{\mu}\ri)}{4\omega_{\G p,e} \omega_{\G p,\mu}}}
\left(1-\frac{|\G p|^{2}}{(\omega_{\G p,e}+m_{e})(\omega_{\G p,\mu}+m_{\mu})}\right)  \, , \\[2mm]
%Y^{11}_{\G p} & = & -\frac{p_3}{\sqrt{4 \om_{\G p,e}\om_{\G p,e}}}
%\lf(\sqrt{\frac{\om_{\G p,\mu}+m_\mu}{\om_{\G p,e}+m_e}}+\sqrt{\frac{\om_{\G p,e}+m_e}{%\om_{\G p,\mu}+m_\mu}}\ri) \, , \\[2mm]
Y^{22}_{\G p} & = & -Y^{11}_{\G p} \ = \ \frac{p_3}{\sqrt{4 \om_{\G p,e}\om_{\G p,\mu}}}
\lf(\sqrt{\frac{\om_{\G p,\mu}+m_\mu}{\om_{\G p,e}+m_e}}+\sqrt{\frac{\om_{\G p,e}+m_e}{\om_{\G p,\mu}+m_\mu}}\ri) \, , \\[2mm]
Y^{12}_{\G p} & = & \lf(Y^{21}_{\G p} \ri)^* \ = \ -\frac{p_1-i p_2}{\sqrt{4 \om_{\G p,e}\om_{\G p,\mu}}}
\lf(\sqrt{\frac{\om_{\G p,\mu}+m_\mu}{\om_{\G p,e}+m_e}}+\sqrt{\frac{\om_{\G p,e}+m_e}{\om_{\G p,\mu}+m_\mu}}\ri) \, .
\eea
%
%
%The first order diagrams are reported in figure \ref{fofeyn}
%
%\begin{figure}[ht]
%\begin{minipage}[b]{0.45\linewidth}
%\feynmandiagram [horizontal=i1 to f1] {
  %i1 [particle=\(\nu_{e} \)] -- [fermion] a -- [fermion] f1 [particle=\(\nu_{\mu} \)],
%};  
%\end{minipage}
%\hspace{0.5cm}
%\begin{minipage}[b]{0.45\linewidth}
%\feynmandiagram [horizontal=i1 to f1] {
  %i1 [particle=\(\nu_{e} \)] -- [fermion] a -- [fermion] f1 [particle=\(\nu_{e} \)],
%}; 
%\feynmandiagram [horizontal=i1 to f2, tree layout] {
%i1 --[draw = none] a,
  %f2 [particle=\(\overline{\nu}_{e} \)] -- [fermion] a -- [fermion] f3 [particle=\(\nu_{\mu}  \)]
%};    
%\end{minipage}
%\caption{First order Feynman diagrams}
%\label{fofeyn}
%\end{figure}
The first non-trivial flavor transition process we consider is
\be
|\nu^r_{\G p,e}\ran \ \rightarrow \ |\nu^s_{\G k,\mu}\ran \, , \qquad |\nu^r_{\G p,\si}\ran \equiv \al^{r\dag}_{\G p,\si}|0\ran \, , 
\ee
whose amplitude reads
\bea \non
{}\hspace{-3mm} \mathcal{A}^{rs}_{e \to \mu}(\G p,\G k, ;t_i,t_f) \ &\approx & \ - i m_{e \mu}\de_{r s} \de_{\G k,\G p} W_{\G p}\, \ \int^{t_f}_{t_i} \!\! \dr t \, e^{i\lf(\om_{\G k,\mu}-\om_{\G p,e}\ri)t} \\[2mm]   
& = & m_{e \mu} \,  \de_{r s} \de_{\G k,\G p}  \, \,  \lf(e^{i\lf(\om_{\G p,\mu}-\om_{\G p,e}\ri)t_f}-e^{i \lf(\om_{\G p,\mu}-\om_{\G p,e}\ri)t_i}\ri)  \frac{W_{\G p}}{\om_{\G k,e}-\om_{\G k,\mu}} \ =  \ \de_{r s} \de_{\G k,\G p}  \,  \tilde{\mathcal{A}}_{e \to \mu}(\G k; t_i,t_f) \, , 
\eea
where
\be
\tilde{\mathcal{A}}_{e \to \mu}(\G p; t_i,t_f) \ = \ \frac{m_{e \mu} \, W_{\G p}}{\om_{\G p,e}-\om_{\G p,\mu}} \, \lf(e^{i\lf(\om_{\G p,\mu}-\om_{\G p,e}\ri)t_f}-e^{i \lf(\om_{\G p,\mu}-\om_{\G p,e}\ri)t_i}\ri) \, . 
\ee
Similarly as in the boson case (see Eq.\eqref{probsum}), the oscillation probability is computed by summing over the final density of states, now involving the sum over the helicities:
\bea \non
\mathcal{P}_{e \to \mu}(\G p;\De t) & = & \sum_{\G k,s} |\mathcal{A}^{rs}_{e \to \mu}(\G p, \G k;t_i,t_f)|^2 \ = \ |\tilde{\mathcal{A}}_{e \to \mu}(\G p, t_i,t_f)|^2 \\[2mm]
& = &  W_{\G p}^2 \, \frac{2 m^2_{e \mu} }{\lf(\om_{\G p,e}-\om_{\G p,\mu}\ri)^2}  \lf[1-\cos\lf[\lf(\om_{\G p,\mu}-\om_{\G p,e}\ri)\De t\ri] \ri]  \, , \qquad \ \De t \equiv t_f-t_i \, .
\eea

Another non-trivial process is the decay
\be
|\nu^r_{\G p,e}\ran \ \rightarrow \ |\nu^{s_1}_{\G k_1,e}\ran |\nu^{s_2}_{\G k_2,\mu}\ran |\overline{\nu}^{s_3}_{\G k_3,e}\ran \, .
\ee
The amplitude explicitly reads
\bea \non
&& \mathcal{A}^{r s_1 s_2 s_3}_{e \to e\overline{e} \mu}(\G p,\G k_1,\G k_2, \G k_3;t_i,t_f)  \approx
 -i \, m_{e \mu} \, \, Y_{\G k_2}^{s_3 s_2} \, \de_{\G k_1, \G p}  \de_{\G k_2, - \G k_3}\, \de_{r s_1}   \int^{t_f}_{t_i} \!\! \dr t \, e^{-i\lf(\om_{\G k_2,\mu}+\om_{\G k_2,e}\ri)t}   \\[2mm] \non
&& = \  -m_{e \mu}  \, \de_{r s_1} \,  \de_{\G k_1, \G p}  \de_{\G k_2, - \G k_3} \,  \lf(e^{-i\lf(\om_{\G k_2,\mu}+\om_{\G k_2,e}\ri)t_f}-e^{-i \lf(\om_{\G k_2,\mu}+\om_{\G k_2,e}\ri)t_i}\ri)  \frac{Y_{\G k_2}^{s_2 s_3}}{\om_{\G k_2,e}+\om_{\G k_3,\mu}} \\
&& =  \ \de_{\G k_1, \G p}  \de_{\G k_2, - \G k_3} \,  \de_{r s_1} \, \tilde{\mathcal{A}}^{s_2 s_3}_{e \to e\overline{\mu}\mu}(\G k_2;t_i,t_f) \, , 
\eea
where
\be
\tilde{\mathcal{A}}^{s_2 s_3}_{e \to e\overline{e} \mu }(\G k;t_i,t_f) \ = \ -\frac{m_{e \mu} \, Y^{s_2 s_3}_{\G k}}{\om_{\G k,e}+\om_{\G k,\mu}} \, \lf(e^{-i\lf(\om_{\G k,\mu}+\om_{\G k,e}\ri)t_f}-e^{-i \lf(\om_{\G k,\mu}+\om_{\G k,e}\ri)t_i}\ri) \, . 
\ee
As done above, we thus find the probability as
\be
\mathcal{P}_{e \to e\overline{e} \mu}(\G p;\De t)  \ = \ \sum_{\G k_1,\G k_2,\G k_3} \sum_{s_1,s_2,s_3} |\mathcal{A}^{r s_1 s_2 s_3}_{e \to e\overline{e} \mu}(\G p,\G k_1,\G k_2, \G k_3;t_i,t_f) |^2  \ = \ \sum_{\G k}\sum_{s_2,s_3}|\tilde{\mathcal{A}}^{s_2 s_3}_{e \to e\overline{e} \mu }(\G k;t_i,t_f)|^2\, .
\ee
In the large-$V$ limit
\be
\mathcal{P}_{e \to e\overline{e} \mu}(\G p;\De t)  \ = \ V \sum_{s_2,s_3} \, \int \!\! \frac{\dr^3 \G k}{(2 \pi)^3} \, \frac{\lf(Y^{s_2 s_3}_\G k\ri)^2 }{\lf(\om_{\G k,e}+\om_{\G k,\mu}\ri)^2}  \sin^2\lf(\frac{\lf(\om_{\G k,\mu}+\om_{\G k,e}\ri)\De t}{2}\ri) \, . 
\ee
This is a divergent contribution (vacuum diagram) and it must be subtracted.

Finally, we have the process
\be
|\nu^r_{\G p,e}\ran \ \rightarrow \ |\nu^{s_1}_{\G k_1,e}\ran |\nu^{s_2}_{\G k_2,e}\ran |\overline{\nu}^{s_3}_{\G k_3,\mu}\ran \, , \qquad \G k_1 \neq \G k_2 \ \lor \ s_1 \neq s_2 \, .
\ee
The amplitude explicitly reads
\bea \mathcal{A}^{r s_1 s_2 s_3}_{e \to e e \overline{\mu}}(\G p,\G k_1,\G k_2, \G k_3;t_i,t_f)
\ =  \ \de_{\G k_1, \G p}  \de_{\G k_2, - \G k_3} \,  \de_{r s_1} \, \tilde{\mathcal{A}}^{s_2 s_3}_{e \to e e\overline{\mu}}(\G k_2;t_i,t_f)-\de_{\G k_2, \G p}  \de_{\G k_1, - \G k_3} \,  \de_{r s_2} \, \tilde{\mathcal{A}}^{s_1 s_3}_{e \to e e\overline{\mu}}(\G k_1;t_i,t_f) \, . 
\eea
where $\tilde{\mathcal{A}}^{s_2 s_3}_{e \to e e \overline{\mu} }(\G k;t_i,t_f)=\tilde{\mathcal{A}}^{s_2 s_3}_{e \to e\overline{e} \mu }(\G k;t_i,t_f)$. Note this correctly goes to zero when $\G k_1=\G k_2$ and $s_1=s_2$.
We thus find the probability as
\bea
&& \mathcal{P}_{e \to e e \overline{\mu}}(\G p;\De t)  \ = \ \ha \sum_{\G k_1,\G k_2,\G k_3} \sum_{s_1,s_2,s_3} |\mathcal{A}^{r s_1 s_2 s_3}_{e \to e e \overline{\mu}}(\G p,\G k_1,\G k_2, \G k_3;t_i,t_f) |^2 \non \\[2mm]
&& =  \sum_{\G k,s_2,s_3} |\tilde{\mathcal{A}}^{ s_2 s_3}_{e \to e e\overline{\mu}}(\G k;t_i,t_f)|^2-\sum_{s_3} |\tilde{\mathcal{A}}^{r s_3}_{e \to e e\overline{\mu}}(\G p;t_i,t_f)|^2\, .
\eea
It is clear we can not simply subtract the first piece involving the sum over momenta: this procedure would give a negative probability. This subtle issue can be solved by remembering that asymptotic states have to be representations of momentum.  
Thus, we must 
isolate the contribution with $\G k= \G p$
\bea
 \mathcal{P}_{e \to e e \overline{\mu}}(\G p;\De t)  & = &  \sum_{\G k \neq \G p,s_2,s_3} |\tilde{\mathcal{A}}^{ s_2 s_3}_{e \to e e\overline{\mu}}(\G k;t_i,t_f)|^2+ \sum_{s_2,s_3} |\tilde{\mathcal{A}}^{ s_2 s_3}_{e \to e e\overline{\mu}}(\G p;t_i,t_f)|^2-\sum_{s_3} |\tilde{\mathcal{A}}^{r s_3}_{e \to e e\overline{\mu}}(\G p;t_i,t_f)|^2\non \\[2mm]
& = &  \sum_{\G k \neq \G p,s_2,s_3} |\tilde{\mathcal{A}}^{ s_2 s_3}_{e \to e e\overline{\mu}}(\G k;t_i,t_f)|^2+\sum_{s_3} |\tilde{\mathcal{A}}^{r s_3}_{e \to e e\overline{\mu}}(\G p;t_i,t_f)|^2 \, .
\eea
In other words, because of the Pauli principle, the vacuum should not carry the contribution with $\boldsymbol{k}=\boldsymbol{p}$.
Then, in the large-$V$ limit
\be
\mathcal{P}_{e \to e e \overline{\mu}}(\G p;\De t)  \ = \ V \sum_{s_2,s_3} \, \int \!\! \frac{\dr^3 \G k}{(2 \pi)^3} \, |\tilde{\mathcal{A}}^{ s_2 s_3}_{e \to e e \overline{\mu}}(\G k;t_i,t_f)|^2+\sum_{s_3} |\tilde{\mathcal{A}}^{r s_3}_{e \to e e\overline{\mu}}(\G p;t_i,t_f)|^2 \, . 
\ee
The first piece diverges and must be subtracted, while the second piece gives a finite contribution. Explicitly
\be
\mathcal{P}_{e \to e e \overline{\mu}}(\G p;\De t)  \ = \  \frac{4 m_{e \mu}^2 Y_\G p^2 }{\lf(\om_{\G p,e}+\om_{\G p,\mu}\ri)^2}  \sin^2\lf(\frac{\lf(\om_{\G p,\mu}+\om_{\G p,e}\ri)\De t}{2}\ri)  \, ,
\ee
where
\be
Y_{\G p}^2 \ = \ \sum_{\G s} \lf(Y^{rs}_{\G p}\ri)^* Y^{rs}_{\G p} \, ,
\ee
and
\be
Y_{\G p} \ = \ \frac{|\G p|}{\sqrt{4 \om_{\G p,e}\om_{\G p,\mu}}}
\lf(\sqrt{\frac{\om_{\G p,\mu}+m_\mu}{\om_{\G p,e}+m_e}}+\sqrt{\frac{\om_{\G p,e}+m_e}{\om_{\G p,\mu}+m_\mu}}\ri) \, .
\ee

Therefore, the total decay probability of $\nu_e$ is
\be
\mathcal{P}^{e}_{D}(\G p; \De t) \ = \ 4 m^2_{e \mu} \lf[ \frac{W_{\G p}^2}{\lf(\om_{\G p,e}-\om_{\G p,\mu}\ri)^2}  \sin^2\lf(\frac{\lf(\om_{\G p,\mu}-\om_{\G p,e}\ri)\De t}{2}\ri)+ \frac{Y_\G p^2 }{\lf(\om_{\G p,e}+\om_{\G p,\mu}\ri)^2}  \sin^2\lf(\frac{\lf(\om_{\G p,\mu}+\om_{\G p,e}\ri)\De t}{2}\ri) \ri] \, . 
\ee
Note that all such decays (flavor transitions) are forbidden when $t_i \to -\infty$, $t_f \to +\infty$, unless $m_e = m_\mu$. In particular, the last two processes are always forbidden for infinite time-intervals because of  energy conservation. In fact, in that case, the three-dimensional $\de$s would be substituted by deltas on the four-momentum, which strictly employ energy conservation. As we commented above, this is expected in analogy of what happens for unstable particles.

 If we now define, following the notation of Ref. \cite{Blasone:1995zc}
\bea
 |U_\G p| & = & W_{\G p}\frac{m_\mu-m_e}{\om_{\G p,e}-\om_{\G p,\mu}} \ = \ \sqrt{\frac{\lf(\om_{\G p,e}+m_e\ri)\lf(\om_{\G p,\mu}+m_\mu\ri)}{4\om_{\G p,e}\om_{\G p,\mu}}}  \lf(1+\frac{|\G p|^2}{\lf(\om_{\G p,e}+m_e\ri)\lf(\om_{\G p,\mu}+m_\mu\ri)}\ri)\, ,\\[2mm]
|V_\G p| & = & Y_{\G p}\frac{m_\mu-m_e}{\om_{\G p,e}+\om_{\G p,\mu}} \ = \ \sqrt{\frac{\lf(\om_{\G p,e}+m_e\ri)\lf(\om_{\G p,\mu}+m_\mu\ri)}{4\om_{\G p,e}\om_{\G p,\mu}}} \lf(\frac{|\G p|}{\om_{\G p,e}+m_e}-\frac{|\G p|}{\om_{\G p,\mu}+m_\mu}\ri)\, ,
\eea
we can rewrite the probability in the form
\be \label{mprob}
\mathcal{P}^{e}_{D}(\G p; \De t)  \ = \ \sin^2 2 \theta \lf[ |U_\G p|^2 \sin^2\lf(\frac{\lf(\om_{\G p,\mu}-\om_{\G p,e}\ri)\De t}{2}\ri)+ |V_\G p|^2  \sin^2\lf(\frac{\lf(\om_{\G p,\mu}+\om_{\G p,e}\ri)\De t}{2}\ri) \ri] \, . 
\ee
with $\theta=m_{e \mu}/(m_\mu-m_e) \approx \sin \theta$.
In the approximation we used this coincides with the oscillation probability \eqref{oscfor}, firstly derived in \cite{BHV99}. This is a remarkable result because the method we adopted is quite different from the approach of Ref. \cite{BHV99} (which is briefly reviewed in appendix), and it does not rely on the construction of a \emph{flavor vacuum} (see Eq.\eqref{flavvac}). 

The second term on the r.h.s. of the Eq.\eqref{mprob} is the main correction with respect to the usual Pontecorvo oscillation formula \eqref{stafor}. The effect of such term is negligible for relativistic neutrinos, e.g. when $m_\si/|\G p| \to 0$, while is maximal when $|\G p| = \sqrt{m_e m_\mu}$ which, in our approximation is equivalent to $|\G p| = \sqrt{m_1 m_2}$, i.e. when the momentum is of the order of neutrino masses. In such a case
\be \label{bognr}
|U_\G p|^2  \ = \ 1-|V_\G p|^2 \ = \ \ha + \frac{\xi}{2} \, , 
\ee
where
\be \label{xi}
\xi \ = 2\ \frac{ \sqrt{m_e m_\mu}}{m_e+m_\mu} \, .
\ee
Possible phenomenological implications in this regime could be studied in a cosmological context, how it will be discussed in the conclusions.

Let us now compute the survival probability $\mathcal{P}^e_{S}(\G k,\De t)$. The zeroth order contribution gives  $\mathcal{P}^e_{S}(\G k,\De t)=1$.  To find a non-trivial contribution to the survival probability we should compute the second order terms. In the present case it is useful to write 
\bea\non
&& U(t_i,t_f) \ = \ \ide -i \, m_{e \mu} \int^{t_f}_{t_i} \!\! \dr^4 x \, : \overline{\nu}_e(x) \nu_\mu(x)+\overline{\nu}_\mu(x) \nu_e(x): \\[2mm]
&& - \  \frac{m^2_{e \mu}}{2} \int^{t_f}_{t_i} \!\! \dr^4 x_1 \int^{t_f}_{t_i} \!\! \dr^4 x_2 \, \, \mathcal{T}\Big[   \lf(:\overline{\nu}_e(x_1) \nu_\mu(x_1)+\overline{\nu}_\mu(x_1) \nu_e(x_1):\ri) \lf( : \overline{\nu}_e(x_2) \nu_\mu(x_2)+\overline{\nu}_\mu(x_2) \nu_e(x_2):\ri)\Big] \, + \, \ldots \, . 
\eea
The second order piece can be further expanded using  Wick theorem:
\bea \non
U^{(2)}(t_i,t_f) & = & -\frac{m^2_{e \mu}}{2} \int^{t_f}_{t_i} \!\! \dr^4 x_1 \int^{t_f}_{t_i} \!\! \dr^4 x_2 \, \,    \Big[:\overline{\nu}_e(x_1) \nu_\mu(x_1)\overline{\nu}_e(x_2) \nu_\mu(x_2):+:\overline{\nu}_e(x_1) \nu_\mu(x_1)\overline{\nu}_\mu(x_2) \nu_e(x_2):   \non \\[2mm]
& + & :\overline{\nu}_\mu(x_1) \nu_e(x_1)\overline{\nu}_e(x_2) \nu_\mu(x_2):+:\overline{\nu}_\mu(x_1) \nu_e(x_1)\overline{\nu}_\mu(x_2) \nu_e(x_2): \non \\[2mm]
&+&  2 \,  i \lf(S^e_{\al \bt}(x_2-x_1) \, :\overline{\nu}^\bt_\mu(x_2) \nu^\al_\mu(x_1):+ S^\mu_{\al \bt}(x_2-x_1) \, :\overline{\nu}^\bt_e(x_2) \nu^\al_e(x_1):\ri) \Big] \, , \label{secu}
\eea
where
\be
S^\si_{\al \bt}(x) \ = \ \int \!\! \frac{\dr^4 p}{(2 \pi)^4} \, e^{-i p x} \, \frac{\lf(\slashed{p}+m_\si\ri)_{\al \bt}}{p^2-m^2_\si+i \varepsilon} \, , \qquad \si=e,\mu \, ,
\ee
is the Dirac propagator. The employment of Wick theorem makes evident that the divergent vacuum contributions come from the terms containing the propagator, which can be neglected.

The survival process is
\be
|\nu^r_{\G p,e}\ran \ \rightarrow \ |\nu^s_{\G k,e}\ran \, .
\ee
Saving only up to linear terms in $m_{e \mu}$, the amplitude can be written as
\be
\mathcal{A}^{rs}_{e \to e}(\G p,\G k; t_i,t_f)=\de_{\G k, \G p} \de_{rs}+\ha \mathcal{A}^{(2) r s}_{e \to e}(\G p,\G k; t_i,t_f) \, , 
\ee
where $A^{(2) r s}_{e \to e}(\G k,\G p;t_i,t_f)$ is the second-order piece, which goes as $m_{e \mu}^2$. Taking the square, and summing over the final momenta and helicities, the only pieces which cannot be disregarded are the one which go as $m_{e \mu}^2$ or lower powers, i.e.
\be
\mathcal{P}^{e}_{S}(\G p; \De t) \ = \ \sum_{\G k,s}\mathcal{A}^{rs}_{e \to e}(\G p,\G k, t_i,t_f) \ \approx \ 1+ 2 \, \Re e \lf(\tilde{\mathcal{A}}^{(2)}_{e \to e}(\G p;t_i,t_f)\ri) \, , 
\ee
with
\be
\tilde{\mathcal{A}}^{(2)}_{e \to e}(\G p;t_i,t_f) \ \equiv \ \sum_{\G k,s} A^{(2) r s}_{e \to e}(\G p,\G k;t_i,t_f) \, .
\ee
Explicitly one finds
\be
\mathcal{P}^{e}_{S}(\G p; \De t) \ = \   1-\sin^2 2 \theta \lf[ |U_\G p|^2 \sin^2\lf(\frac{\lf(\om_{\G p,\mu}-\om_{\G p,e}\ri)\De t}{2}\ri)+ |V_\G p|^2  \sin^2\lf(\frac{\lf(\om_{\G p,\mu}+\om_{\G p,e}\ri)\De t}{2}\ri) \ri] \, . 
\ee
Then
\be
\mathcal{P}^{e}_{D}(\G p; \De t) + \mathcal{P}^{e}_{S}(\G p; \De t)\ = \ 1 \, ,
\ee
as expected.
%%%%%%%%%%%%%%%%%%%%%%%%%%%%%%%%%%%%%%%%%%%%%%%%%%%%%
\section{Conclusions} \label{con}
In this paper we have shown that neutrino oscillations can actually be described as a standard perturbative  QFT, with  the mixing handled by the interaction picture.
A 0+1D toy model, ``scalar'' neutrinos and their ``realistic'' Dirac fermion counterparts can all be described by this approach, the various transition amplitudes can be related via Feynman diagrams and the differences reduced to spin-statistics differences.   In the fermionic case, the same oscillation formula as in non-perturbative flavor-Fock space approach \cite{BHV99}, is here independently recovered, within the approximations involved in the perturbative calculation. In particular, we find a term which depends on the sum of the frequencies, in addition to the usual Pontecorvo term which only involves their difference. Moreover, the pre-factors in front of such terms are exactly the Bogoliubov coefficients derived in Ref.\cite{Blasone:1995zc}.

This description sheds new light on the long-standing arguments in the literature as to the ``true basis'' of neutrino Fock space.   As in other quantum field theories, provided interacting and free states have the same quantum numbers, the Kallen-Lehmann representation can be used to convert between the two perturbatively. In such a picture, experimental data selects the physical degrees of freedom, to be related to the Lagrangian ones via Renormalization.   Since interaction happens via weak charge, the perturbation series defined in this work shows that neutrino physical states can be consistently defined.

The methods developed in this work can be applied to a  calculation of observables beyond tree level which systematically includes oscillation effects.  An obvious candidate is the magnetic dipole moment of the neutrino, which so far has only been obtained by effective field theory techniques \cite{shrock}.    A proper calculation will include the interaction included here (which corrects the magnetic mass) with the same footing as the W-lepton bubble (which provides the electric current generating the magnetic moment).  Given the non-trivial momentum dependence of QFT neutrino oscillation effects noticed in \cite{Blasone:1995zc} the interplay of all these effects could yield surprises, which might be observable \cite{yago}.
  
There are however subtleties not present in other usual quantum field theories.   Unlike other interactions, neutrino oscillations are 2-field operators, analogous to coupled harmonic oscillators.    The vacuum of the theory should therefore be analytically obtainable, and it is well known that perturbative terms can not capture such vacuum corrections.     As a related issue, the self-interaction due to the vacuum self energy will result in a ``mass'' correction to the flavor state, tilting the perturbative series examined in this work.
Since the theory examined here is perfectly Lorentz invariant, violations derived using the quantum field theory approach to fermion mixing (see e.g. \cite{Blasone:2004yh}) can only be a feature of the vacuum and to properly assess such effects a perturbative series must be derived from such a vacuum.   This is left to a forthcoming work.

Moreover, it is interesting to stress that the physical corrections to the usual neutrino oscillation formula \eqref{stafor} could be relevant in the measurements of the so-called \emph{cosmic neutrino background} (CNB). This will be studied, e.g., by the PTOLEMY experiment \cite{Betts:2013uya,Cocco:2017nax,Messina:2018xzi,PTOLEMY:2019hkd}. The basic idea of such experiment is to detect the relic neutrinos of the CNB (which decoupled around one second after the Big Bang) through the neutrino capture by tritium $\nu_e +{}^3 {\rm H}\to e^- + {}^3{\rm He}$. With the standard approach to neutrino oscillations, the rate of the capture process reads \cite{PTOLEMY:2019hkd}
\be
\Ga_{CNB} \ = \ \sum_j \, |U_{e j}|^2 \, \bar{\si} \, v_\nu \, f_{e,i} \, n_0 \, , 
\ee
where $U$ is the mixing matrix, $v_\nu$ is the neutrino velocity  as measured at Earth, $\bar{\si}$ is the average cross-section, $n_0$ is the average neutrino number density on large scale and $f_{e,i}$ are the clustering factors. We expect that the analysis of this paper modifies the above expression, with the inclusion of the coefficients $|U_\G p|$ and $|V_\G p|$. This could be easily understood looking at Ref. \cite{Lee:2017cqf}, where the rate of $\beta$-decay ${}^3 {\rm H} \to e^- + {}^3{\rm He}+\nu_e$ was computed within the flavor Fock-space approach.
As discussed in the text (see Eqs. \eqref{bognr}-\eqref{xi}), the corrections in that case, and then in the present case, become  relevant in the non-relativistic regime. However, a quantitative analysis requires to consider the three-flavor phenomenology and it will be postponed to a forthcoming work.

Finally, another phenomenon which is suited to be investigated within the approach here introduced are {\em chiral oscillations} \cite{Bernardini:2005wh,Bittencourt:2020xen,Bittencourt:2022hwn}, which again are relevant in the non-relativistic regime and have been recently discussed in connection with CNB \cite{Ge:2020aen}. Work is in progress in this direction.

%%%%%%%%%%%%%%%%%%%%%%%%%%% \%%%%%%%%%%%%%%%%%%%%%%%%%%
\section*{Acknowledgments}
%{\bf The authors thank the anonymous Referee for the the advices which helped to improve the quality and the interest of the manuscript.}
L.S. was supported by the Polish National Science Center grant 2018/31/D/ST2/02048.
G.T.~acknowledges support from Bolsa de produtividade CNPQ
306152/2020-7, Bolsa de pesquisa FAPESP 2021/01700-2, Partecipation
to Tematico FAPESP, 2017/05685-2 and the
grant BPN/ULM/2021/1/00039 from the Polish National Agency for Academic Exchange.
F. G. thanks S. Mr\'{o}wczy\'{n}ski for useful discussions and financial support from the Polish National Science Centre (NCN) via the OPUS project
2019/ 33/B/ST2/00613.
%%%%%%%%%%%%%%%%%%%%%%%%%%%%%%%%%%%%%%%%%%%%%%%%%%%%%%%%%%%%%%%%%%%%%
\appendix
%%%%%%%%%%%%%%%%%%%%%%%%%%%%%%%%%%%%%%%%%%%%%%%%%%%%%%%%%%%%%%%%%%%%%%%%%%%%%%%%%%%%%%%%%%%%%%%%%%%%%
\section{The non-perturbative approach -- the Heisenberg picture} \label{comp}
The mixing transformation \eqref{mixtra} can be rewritten as \cite{Blasone:1995zc}
\bea
\label{MixingRel1}
\nu_\si(x)=
G_{\theta}^{-1}(t)\, \nu_j (x) \,  G_{\theta}(t) \, , \label{MixingRel2}
\eea
with $(\si,j)=(e,1),(\mu,2)$ and  $G_\theta(t)$  given by
\be
\mbox{\hspace{-2mm}}G_{\theta}(t)=\exp\left[\theta\intx \, \left(\nu_1^{\dagger}(x)\nu_2(x)-
\nu_2^{\dagger}(x)\nu_1(x)\right)\right].
\label{MixGen}
\ee
Fields with definite masses can be expanded as
\begin{eqnarray}
\nu_{j}(x) =  \sum_r \,  \int \!\! \frac{\dr^3 k}{(2 \pi)^{\frac{3}{2}}} \,  \left[ u_{{\bf k},j}^{r}(t) \, \alpha_{{\bf k},j}^{r} + v_{-{\bf k},j}^{r}(t) \, \bt_{-{\bf k},j}^{r\dag}   \right]  e^{i{\bf k}\cdot {\bf x}}  \, ,
\label{fieldex2}
\end{eqnarray}
with $u^r_{{\bf k},j}(t) \,= \, e^{- i \om_{\G k,j} t}\, u^r_{{\bf k},j}\;$,
$\;v^r_{{\bf k},j}(t) \,= \, e^{ i \om_{\G k,j} t}\, v^r_{{\bf k},j}$,
 $\om_{\G k,j}=\sqrt{|\G k|^2 + m_j^2}$.
From (\ref{fieldex2}) and (\ref{MixingRel1}) it follows that flavor fields can be also Fourier expanded:
\begin{eqnarray}
\nu _{\si}(x) \ = \ \sum_r \,  \int \!\! \frac{\dr^3 k}{(2 \pi)^{\frac{3}{2}}}\!\! \ \,
\left[ u_{{\bf k},j}^{r}(t) \, \alpha _{{\bf
k},\si}^{r}(t) \, + \, v_{-{\bf k},j}^{r}(t) \, \beta _{-{\bf k},\si}^{r\dagger
}(t)\right] \, e^{i{\bf k}\cdot {\bf x}} \, ,
\label{fieldex1}
\end{eqnarray}
where the flavor ladder operators are given by
\bea\label{flava}
\begin{pmatrix}
\alpha^{r}_{{\bf
k},\si}(t)\\
\beta^{r}_{-{\bf k},\si}(t)
\end{pmatrix}
\ = \
G_{\theta}^{-1}(t)\,
\begin{pmatrix}
\alpha^{r}_{{\bf
k},j}(t)\\
\beta^{r}_{-{\bf k},j}(t)
\end{pmatrix}
\,  G_{\theta}(t) \, .
\eea
In the Heisenberg picture, the \emph{flavor vacuum} is
\be 
\label{flavvac} |0\rangle_{e,\mu} = G^{-1}_{\theta}(0)\;
|0 \rangle_{1,2}\; , 
\ee
where $|0\ran_{1,2}$ denotes the \emph{mass vacuum}, annihilated by $\al^r_{\G k,j}$,$\bt^r_{\G k,j}$. 
One can easily verify that $|0\rangle_{e,\mu}$ is annihilated by the flavor operators
introduced in Eq.~\eqref{flava}. 
Now, flavor states can be naturally defined as 
\be \label{bvflavstate}
|\nu^r_{\G k,\si}\ran \ = \ \al^{r\dag}_{\G k,\si} |0\ran_{e ,\mu}  \, .
\ee
where flavor operators are taken at reference time $t=0$. These are eigenstates of the lepton/flavor charges
\be
Q_{\si}(t)  \ \equiv \ \intx \, \nu^\dag_\si(x) \, \nu_\si(x) \, ,
\ee
i.e.
\be
Q_{\si}(0) |\nu^r_{\G k,\si}\ran \ = \ |\nu^r_{\G k,\si}\ran \, .
\ee

We can then evaluate oscillation formulas as the expectation value of the flavor charges on a reference neutrino state \cite{BHV99}
\be
\mathcal{Q}_{\si\rightarrow \rho}(t) \ = \ \lan Q_{\rho}(t) \ran_\si \, ,
\ee
where $\langle \cdots\rangle_\si = \lan \nu^r_{\G k,\si}| \cdots |\nu^r_{\G k,\si}\ran$. Explicitly
\bea \label{oscfor}
&& \mbox{\hspace{-4mm}}\mathcal{Q}_{\si\rightarrow \rho}(t)  =   \sin^2 (2 \theta)\Big[|U_\G k|^2\sin^2\lf(\frac{\om_{\G k}^{_-}}{2}t\ri)+  |V_\G k|^2\sin^2\lf(\frac{\om_{\G k}^{_+}}{2}t\ri)\Big]  , \nonumber \\[1mm]
&& \mbox{\hspace{-4mm}}\mathcal{Q}_{\si\rightarrow \si}(t)  =  1 \ - \ \mathcal{Q}_{\si\rightarrow \rho}(t) \, , \quad \si \neq \rho \, ,
\eea
with $\om_{\G k}^{_{\pm}}\equiv \om_{\G k,2}\pm\om_{\G k,1}$, and 
\begin{eqnarray}
U_{{\bf k}}(t)& \equiv & u^{r\dag}_{{\bf k},2}u^r_{{\bf k},1}\;
e^{i(\om_{\G k,2}-\om_{\G k,1})t} \ = \ |U_\G k| \, e^{i(\om_{\G k,2}-\om_{\G k,1})t} \,  \, ,  \\[2mm]
V_{{\bf k}}(t) & \equiv & \epsilon^r\; u^{r\dag}_{{\bf k},1}v^r_{-{\bf k},2}\;
e^{i(\om_{\G k,2}+\om_{\G k,1})t} \ = \ |V_\G k| \, e^{i(\om_{\G k,2}+\om_{\G k,1})t} \, .
\end{eqnarray}
Explicitly
\bea
&& |U_\G k| \ = \ \mathcal{A}_\G k  \lf(1+\frac{|\G k|^2}{\lf(\om_{\G k,1}+m_1\ri)\lf(\om_{\G k,2}+m_2\ri)}\ri)\, ,\qquad
|V_\G k| \ = \ \mathcal{A}_\G k \lf(\frac{|\G k|}{\om_{\G k,1}+m_1}-\frac{|\G k|}{\om_{\G k,2}+m_2}\ri)\, ,
\eea
with $\mathcal{A}_\G k =  \sqrt{\frac{\lf(\om_{\G k,1}+m_1\ri)\lf(\om_{\G k,2}+m_2\ri)}{4\om_{\G k,1}\om_{\G k,2}}}$,  $|U_\G k|^2=1-|V_\G k|^2$. Note that
\be
 \mathcal{Q}_{\si\rightarrow \rho}(t) \ \approx \ \mathcal{P}^{(P)}_{\si\rightarrow \rho}(t) \qquad \mathrm{when} \qquad m_i/|\G k| \to 0 \, , \, \om^{-}_\G k \ \neq \ 0 \, ,
\ee
i.e. in the relativistic limit one re-obtained the Pontecorvo formula \cite{Bilenky:1977ne}
\be 	\label{stafor}
\mathcal{P}^{(P)}_{e\rightarrow \mu}(t) \ = \ \sin^2 (2 \theta)\,\sin^2 \lf(\frac{\om_{\G k,1}-\om_{\G k,2}}{2}t\ri) \, .
\ee
We thus recover the usual phenomenological results in such limit. Notice that formulas \eqref{oscfor} are also recovered in a relativistic QM treatment based on Dirac equation \cite{Bernardini:2005wh}.

%%%%%%%%%%%%%%%%%%%%%%%%%%%%%%%%%%%%%%%%%%%%%%%%%%%%%%%%%%%%%%%%%%%%%%%
%%%%%%%%%%%%%%%%%%%%%%%%%%%%%%%%%%%%%%%%%%%%%%%%%%%%%%%%%%%%%%%%%%%%%%%%
\bibliography{LibraryNeutrino}

\bibliographystyle{apsrev4-2}

\end{document}